\documentclass[epj,twocolumn]{webofc}
\usepackage[varg]{txfonts}   
\usepackage{graphicx}

\newcommand{\Msun}{$M_{\odot}$}

\wocname{epj}
\woctitle{Seismology of the Sun and the Distant Stars 2016}
\begin{document}
\title{The potential of space observations for pulsating pre-main sequence stars}
%

\author{\firstname{Konstanze} \lastname{Zwintz}\inst{1}\fnsep\thanks{\email{konstanze.zwintz@uibk.ac.at}}
}

\institute{Universit\"at Innsbruck, Institute for Astro- and Particle Physics, Technikerstrasse 25, A-6020 Innsbruck 
          }

\abstract{%
The first asteroseismic studies of pre-main sequence (pre-MS) pulsators have been conducted based on data from the space telescopes MOST and CoRoT with typical time bases of less than 40 days. With these data, a relation between the pulsational properties of pre-MS $\delta$ Scuti stars and their relative evolutionary phase on their way from the birthline to the zero-age main sequence was revealed. 
But it is evident from comparison with the more evolved pulsators in their main sequence or post-main sequence stages observed by the main Kepler mission, that many more questions could be addressed with significantly longer time bases and ultra-high precision. 

Here, I will discuss the observational status of pre-MS asteroseismology and the potential of future space observations for this research field.
}
\maketitle
%
\section{Introduction}
\label{intro}

A wealth of different types of pulsators has been studied extensively with data from the Kepler space telescope \cite{koc10} and brought new insights into the physical processes acting inside these stars. As the field selected for the main Kepler mission pointed away from young (star forming) regions, we are lacking high-precision Kepler photometry for pulsating pre-main sequence (pre-MS) objects. Using space data obtained with the MOST \cite{wal03} and CoRoT \cite{bag06} satellites for a maximum of 39 days and the first Kepler K2 data \cite{how14} for a maximum of 80 days, we can illustrate the potential of space observations for the pre-MS stages of pulsating stars and the limitations of our current observational material.

\section{Pre-main sequence evolution}
\label{sec-prems}

\begin{figure*}[htb]
\centering
\includegraphics[width=\hsize,clip]{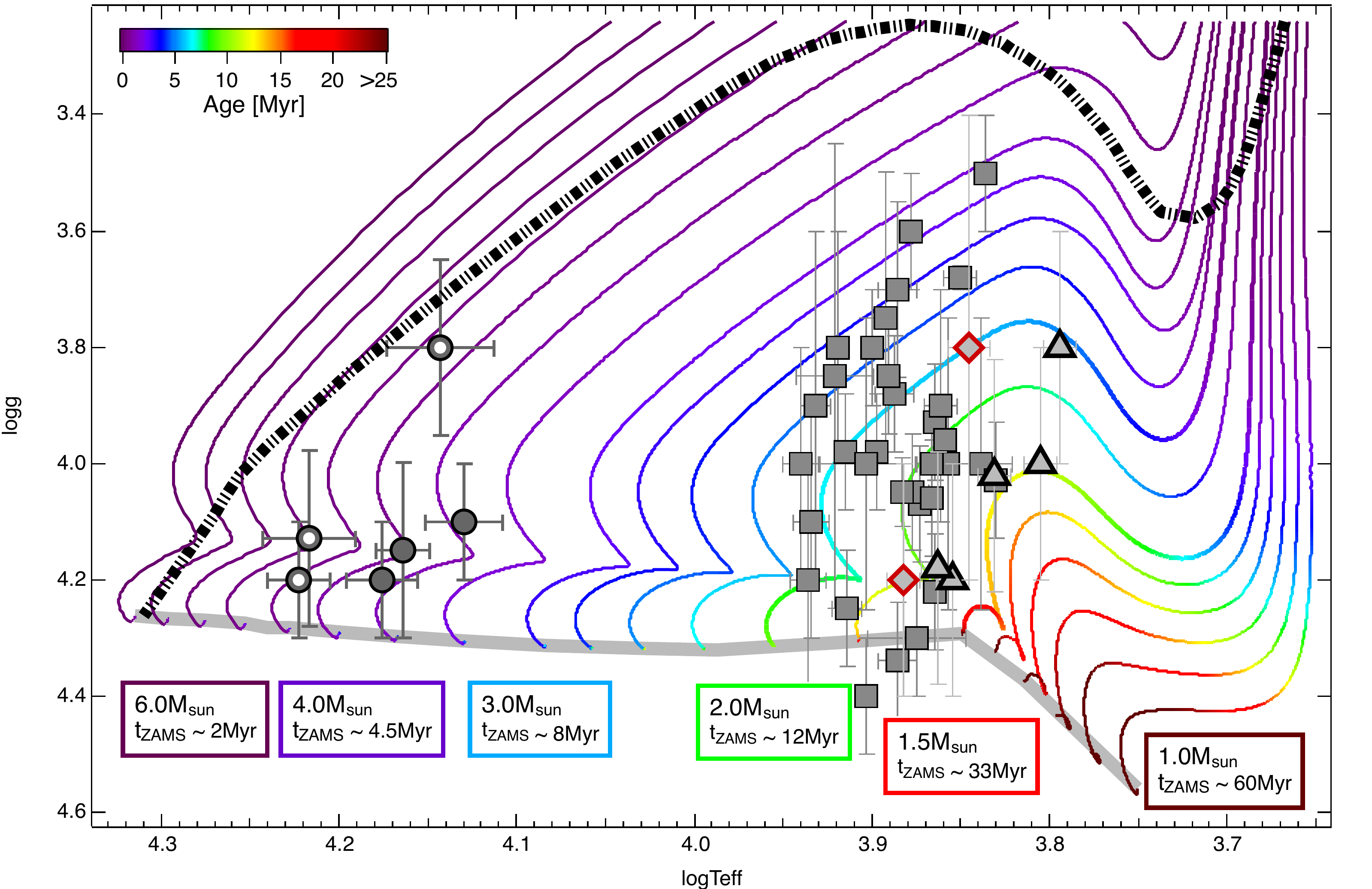}
\caption{Kiel Diagram showing the currently known pre-MS pulsators. Evolutionary tracks from 1 to 6\,\Msun\,\, are taken from Guenther et al. \cite{gue09} and are color coded with age. The birthline is given as thick dash-dotted line, while the ZAMS is marked in light grey. Pre-MS $\delta$ Scuti stars are illustrated as grey squares, pre-MS $\gamma$ Doradus stars as grey triangles, pre-MS $\delta$ Scuti -- $\gamma$ Doradus hybrid stars as grey diamonds with red borders and potential pre-MS SPB type stars as grey circles where stars in binary systems are additionally marked with a white dot.}
\label{hrd_evol}
\end{figure*}

Pre-MS stars do not have started nuclear fusions yet. They mainly gain their energy from gravitational contraction, and through this they get more compact and hotter on their way from the birthline to the zero-age main sequence (ZAMS). 
The speed of evolution in the pre-MS phase is quite rapid which is illustrated in Figure \ref{hrd_evol}. According to our current models, a 1\,\Msun\, star needs $\sim$60 Myrs to reach the ZAMS. It takes only $\sim$12 Myr for a 2\,\Msun\, star, and $\sim$4.5 Myr for a 4\,\Msun\, star. At a mass of around 6\,\Msun\, the birthline intersects the ZAMS, hence more massive stars do not have an optically visible pre-MS phase.
Of course, here we are under several assumptions: first we assume our models are correct - but we know that we are lacking several physical ingredients and a better understanding of some quite crucial processes such as the influence of accretion from the birth environment, initial chemistry (on the surface and inside stars), convection that dominates young stars, magnetic fields or rotation. The evolutionary tracks used in Figure \ref{hrd_evol}, for example, are calculated without rotation. Inclusion of rotation will shift the tracks significantly. 

One way to approach answering some of these open issues, is using asteroseismology.
To do so, we need to populate these regions with objects that are in their pre-MS phase and show pulsations.

\section{Pulsating pre-main sequence stars}

\subsection{$\delta$ Scuti type}


The firstly discovered and hence largest group of pre-MS pulsators are the $\delta$ Scuti type objects that span the mass range between about 1.5 and 3.5 \Msun.
The group of objects that we can use for the first detailed studies comprises of 34 objects that were observed photometrically from space by MOST \cite{wal03} and CoRoT \cite{bag06} and one object recently also by K2 \cite{how14}. 
To put these stars into a Kiel diagram, their fundamental parameters and chemical abundances were derived from dedicated high-resolution spectroscopic observations obtained with the McDonald Observatory 2.7m telescope (US), the ESO VLT (Chile), the Canada France Hawaii Telescope (CFHT; Hawaii, US) and the 1.2m Mercator telescope (La Palma, Spain).
Using the pulsation frequencies from space photometry and the fundamental parameters from spectroscopy, we found an observational relation between the highest excited $p$-mode frequency, named $f_{\rm max}$, and the evolutionary stage (see Figure 1 in \cite{zwi14}). Obviously such a relation could be the basis of an empirical scaling relation - similar as we know it for the solar-like oscillators.

\begin{figure}[htb]
\centering
\includegraphics[width=\hsize,clip]{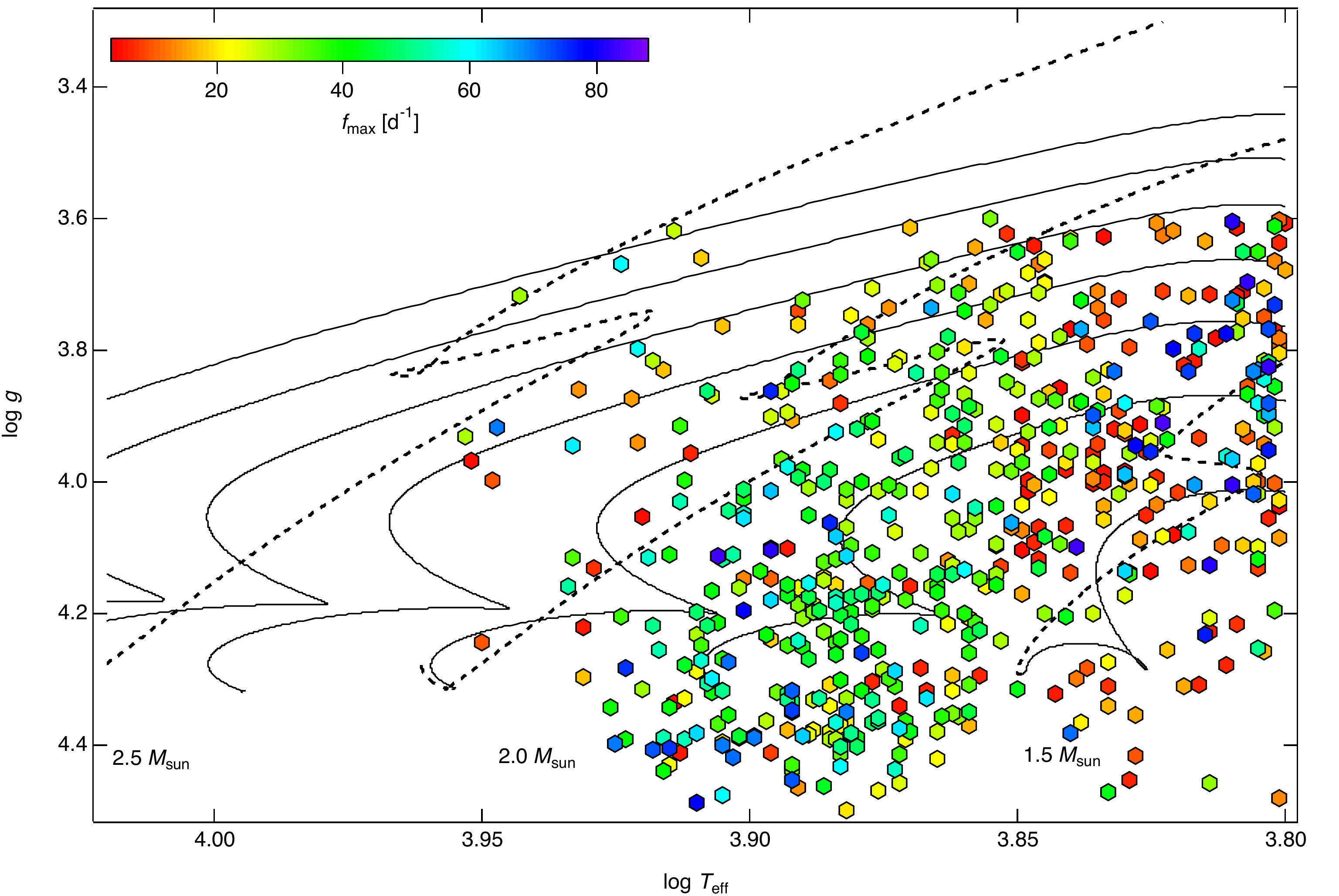}
\caption{Kiel diagram showing $\sim$1000 classical main sequence and post-main sequence $\delta$ Scuti stars observed by CoRoT in the IR and LRa01 runs with spectroscopic observations from the Australian Astronomical Observatory \cite{kai14}. The symbols are color-coded with the highest excited $p$-mode frequency, $f_{\rm max}$. A relation like in the case for pre-MS $\delta$ Scuti stars cannot be seen.}
\label{postms-dscts}
\end{figure}

For the classical post-MS $\delta$ Scuti pulsators this relation cannot be found. Figure \ref{postms-dscts} shows roughly 1000 $\delta$ Scuti stars observed by CoRoT in the IR and LRa01 for which spectroscopic observations were available from the Australian Astronomical Observatory \cite{kai14}. A relation as described before is not seen.

The next observational step for pre-MS $\delta$ Scuti stars will be from the Kepler K2 campaign 9 data. 

\subsection{$\gamma$ Doradus and $\delta$ Scuti - $\gamma$ Doradus hybrid types}

Figure \ref{hrd_evol} shows that not many $\gamma$ Doradus and $\delta$ Scuti - $\gamma$ Doradus hybrid stars have been identified yet: Using CoRoT space photometry, Ripepi et a. \cite{rip11} found one hybrid pre-MS pulsator, and Zwintz et al. \cite{zwi13} discovered two pre-MS $\gamma$ Doradus stars. At least four more objects are currently under investigation (Zwintz et al., in preparation). This number is yet insufficient to conduct a study comparable to the one done for pre-MS $\delta$ Scuti stars \cite{zwi14}. Therefore, the first and immediate potential for future space missions is to discover more bona-fide members of the group of pre-MS $\gamma$ Doradus stars to be able constrain their asteroseismic group properties.

Secondly, the positions of the borders of the instability region for $\gamma$ Doradus pulsators in the HR diagram depend on how convection is treated in the theoretical models, and, hence, is quite uncertain \cite{bou11}. With a statistically significant number of (pre-MS) $\gamma$ Doradus stars whose location in the Kiel diagram is derived from high-resolution spectroscopy, we might be able to give an observational input to theory and put constraints on our theoretical concept for convection.

The ultimate goal for an observational study of pre-MS $\gamma$ Doradus stars is the detection of prograde and retrograde period spacing patterns as they were discovered and analyzed in the hydrogen burning counterparts, e.g., in KIC 8375138 \cite{van15}. 
We would then be able to see if we find dips like for the (post-) main sequence objects, if we can identify signatures of mixing and rotation and really measure the internal differences of stars in both evolutionary stages that have similar atmospheric properties. 
We can only reach that by a combination of an excellent instrument in space, a suitable candidate target and a long enough time base. 

BRITE-Constellation \cite{wei14} is able to pick up period spacing patterns in a ``classical'', hydrogen core burning $\gamma$ Doradus pulsator with photometric time series obtained within 156 days. This tells us that light curves of approximately that length are needed for pre-MS $\gamma$ Doradus stars to clearly pick up period spacing patterns. Using 39-days light curves obtained with CoRoT on stars in NGC 2264 we detect first signs of period spacings in pre-MS $\gamma$ Doradus stars (Zwintz et al., in prep.). But it is evident that we are missing several periods for a complete spacing sequence due to the insufficient time base of the observations.

\subsection{Slowly Pulsating B (SPB) types}
Pre-MS SPB stars are in the evolutionary transition phase from gravitational contraction to hydrogen core burning as their main energy source. Asteroseismic studies of pre-MS SPB stars would allow to measure the changes of their interior structures right before arrival on the ZAMS. 

Observationally, pre-MS SPB stars are hard to find due to several reasons: (i) As their optically visible pre-MS phase is rather short lasting less than 3.6 million years for a 4 \Msun\,\,star and less than 1.6 Myrs for a 6 \Msun\,\,star, they are statistically relatively rare objects, (ii) a high percentage of B stars are in binary systems \cite{san12} which needs to be taken into account during the analysis, and (iii) rotation plays an important factor in B stars \cite{mey00} which can complicate their analysis. 

Currently only two candidates for pre-MS SPB pulsation have been reported \cite{gru12} and four more candidates are identified in the young cluster NGC 2264 from MOST photometry (Zwintz et al. 2016, submitted). These six objects are marked as filled circles in Figure \ref{hrd_evol}, where the pulsators identified as binaries from spectroscopy are identified with a white dot on top of the grey circle.
For three of these objects first signs of period spacings are detected in photometric time series only $\sim$ 22 and $\sim$ 39 days long. Similar as in the case for pre-MS $\gamma$ Doradus $g$-mode pulsators, the main aim is to find period spacings that lead to a better constraint on rotation and diffusive mixing in young stars as it was done already for main sequence stars (e.g., for KIC 7760680) \cite{pap14,mor16}.

\subsection{Solar like oscillators}
The presence of solar like oscillations in pre-MS objects has been predicted by Samadi et al. 2005 \cite{sam05}. The authors calculate the rate $P$ at which the energy is injected into solar $p$-modes and determine that the maximum rate, $P_{\rm max}$, scales with $(L/M)^{2.6}$ for main sequence stars, but follows a $(L/M)^{3.3}$ relation in the pre-MS phase, where $L$ is the stellar luminosity and $M$ is the stellar mass. Similar as in main sequence stars, also for solar-like pre-MS objects the $p$ mode excitation occurs closer to the surface of the stars which is characterized by the effective temperature, $T_{\rm eff}$, and gravity, log\,$g$. These parameters then relate as $T_{\rm eff}^4 / g$ to $L/M$. 
The authors describe that the difference in their computations lies in the usage of 3D simulations for the main sequence stars and mixing length theory (MLT; \cite{boe58}) for the pre-MS phase, but they also note ``However, the physical reasons [for the different relations] are not yet identified." \cite{sam05}. 

Samadi et al. \cite{sam05} also predict the maximum velocity amplitudes, $v_{\rm max}$ to lie between $\sim$ 20 cm$s^{-1}$ and 120 cm$s^{-1}$ and the luminosity amplitudes, $(dL/L)_{\rm max}$ between $\sim$ 2 and 15 parts-per-million (ppm). This is yet again another observational challenge that might be tackled with future space missions like TESS \cite{ric14} and Plato \cite{rau14} because pre-MS solar-like oscillators are located in the same region of the Hertzsprung-Russel diagram (HRD) as the T Tauri stars and they also share the early evolutionary stage. T Tauri stars are characterized by a high degree of activity and show regular and irregular light variations of a magnitude and more (see, e.g., \cite{ale10, cod14}). Searching for pulsation amplitudes of a few ppm in stars with such a high level of activity requires the longest possible photometric time series obtainable from space. A first search for solar-like variability in the CoRoT NGC 2264 observations using a time base of 39 days did not reveal any signature of stochastic oscillations.

\section{Conclusions}
\label{sec-con}
The main potential of future space photometry for pre-MS pulsators lies in a crucial increase of the photometric time bases from typically 40 days (with up to 80 days maximum from K2) to at least 150 days and more. With shorter light curves we will be able to identify new pre-MS $\delta$ Scuti, $\gamma$ Doradus and SPB stars, but only with sufficiently long time series we will be able to detect long sequences of period spacing patterns for the $g$-mode pulsators.
The discovery of solar-like oscillations in pre-MS stars -- as they are predicted by theory -- will also only be possible with long time base measurements which decrease the noise level in the Fourier domain sufficiently enough to detect signal at the ppm level.

Using the available data from MOST and CoRoT for pre-MS $\delta$ Scuti stars, we have first independent proof from asteroseismology that stars are formed sequentially in clusters \cite{zwi14}, that giving one age for all cluster members is not sufficient and will always be affected by large errors in particular for the youngest objects. 
With new ultra-precise observations from space and time bases of 150 days and more, we would be able to seismically calibrate our pre-MS stellar structure models, and, hence, contribute to a better understanding of early stellar evolution.

The formation of stars is directly connected to the formation of planets. In their recent paper, David et al. \cite{dav16} use K2 data from Upper Sco to study a  Neptune-sized transiting planet closely orbiting a 5-10 million year old star. It is not unlikely that in our future space observations we will find a (young) pulsating star that hosts a planetary system. We will then be able to use asteroseismology to put constraints on the exoplanet host star. The first candidate for this analysis is the young ZAMS star $\beta$ Pictoris which shows $\delta$ Scuti pulsations \cite{koe03} and whose planet will cross the star's Hill sphere in 2017.   

These examples illustrate that pre-MS asteroseismology also has a significant impact on other fields of stellar astrophysics, and that future space missions should put their focus also on this research area.

\begin{acknowledgement}
K.Z. acknowledges support by the Austrian Fonds zur F\"orderung der wissenschaftlichen Forschung (FWF, project V431-NBL).
\end{acknowledgement}

%
%

\end{document}